\documentclass[12pt]{article}

\usepackage{amsmath}
\usepackage{graphicx}

\newcommand{\bra}[1]{\left\langle{}#1\right|}
\newcommand{\cfield}{\mathbf{C}}

\newcommand{\drm}{\,\rm d}
\newcommand{\dst}{\mu}
\newcommand{\ee}{\mathbf{e}}

\newcommand{\entrp}{S}
\newcommand{\hh}{\mathcal{H}}
\newcommand{\ket}[1]{\left| #1\right\rangle}

\newcommand{\lmu}{\Lambda}
\newcommand{\lth}{n}
\newcommand{\mesb}{\,\rmd\psi}

\newcommand{\rfield}{\mathbf{R}}
\newcommand{\rmd}{\rm{d}}
\newcommand{\rme}{\rm{e}}
\newcommand{\rrh}{\rho}
\newcommand{\tpar}{B}

\newcommand{\trm}{\rm Tr \,}

\newcommand{\bracket}[3]{\bra{#1}#2\ket{#3}}
\newcommand{\braket}[2]{\left\langle{}#1\right|\left.#2\right\rangle}
\newcommand{\cbn}{\cfield{B}_\lth}
\newcommand{\ketbra}[2]{\ket{#1}\!\!\bra{#2}}

\newcommand{\prjo}[1]{\ketbra{#1}{#1}}
\newcommand{\raypr}[1]{\ketbra{#1}{#1}}

\begin{document}

\title{Guessing Quantum Ensemble Using Laplace Principle}

\author{Georges Parfionov\thanks{Dept. of Mathematics, SPb EF University,
Griboyedova 30--32, 191023, St.Petersburg, Russia}  and
        Rom\`an R. Zapatrin\thanks{Dept. of Informatics, The State Russian
Museum, In\.{e}nernaya, 4, 191186, St.Petersburg, Russia
(corresponding author, email: Roman.Zapatrin at gmail.com)}}


\maketitle

\begin{abstract}
For a mixed quantum state with density matrix $\rrh$ there are
infinitely many ensembles of pure quantum states, which average to
$\rrh$. Starting from Laplace principle of insufficient reason (not
to give \emph{a priori} preference to any particular state), we
derive a `natural' distribution of pure states averaging to $\rrh$,
which is `more spread' than all the others.
\end{abstract}

\section{Introduction}
\label{intro} In classical situation an unknown probability
distribution can be estimated by collecting statistics. This is
not the case in quantum mechanics when we need to estimate a
distribution of pure quantum states. All we can do is to estimate
the density matrix $\rrh$ of appropriate mixed quantum state.
Although, there are infinitely many distributions of pure quantum
states which average to the density matrix $\rrh$. That is why in
order to estimate the distribution, that is, the ensemble of pure
quantum states, some \emph{a priori} assumption about this
distribution is required. Which one?

It was Laplace who introduced the formula of classical probability
\cite{laplace}
\[
P=\frac{m}{n}
\]
where $n$ stands for the total number of outcomes and $m$ is the
number of favorable ones. This formula is not \emph{ad hoc}
introduced. Rather, it based on the \emph{principle of
insufficient reason}: 
\begin{quote}
if there is no reason to prefer one outcome
of an experiment with respect to another one, all outcomes are
treated equally probable.
\end{quote}

\noindent According to Laplace, if we are given a completely unknown
distribution and we need to estimate it, we assume it to be just
uniform. But what should we do if we have an additional
information about the distribution? Can we still use Laplace
principle?

\section{A classical example: exploring biased die}\label{sclass}

Suppose we play with die whose properties are not known. If we are
asked what is the probability of a certain face to appear, we
intuitively (but in fact according to Laplace) answer: `there are
6 faces, none of them is preferred, hence any face appears with
the same probability $1/6$'.

We roll $N$ identical dice and, as a result of this experiment, we
learn the mean value, denote it $M$, of the number of shown points.
This is an information about the die, how it affects our estimation?
In this case the hypothesis of the equality of all \emph{faces} may
no longer be compatible with initial hypothesis that all faces are
the same (indeed, one would scarcely believe a die showing $M=5$
points at average to be symmetric). So, the Laplace principle is not
applicable, at least in its direct form.

Consider it in a more general setting. Suppose we have $N$
identical dice with $K$ faces each. Each face $k$ is labeled by a
value $A_k$ and the die might be `biased', that is, the
probability of $k$-th face to appear is an unknown number $p_k$.
$N$ such identical dice are rolled, and the average value of the
number appeared turns out to be $M$. What can we now say about
$p_k$?

This average value $M$ can be obtained when we have $n_1$ times face
$1$,\ldots, $n_K$ times face $K$, with the values $\{n_1,\ldots ,
n_K\}$ satisfying the equations
\begin{equation}\label{emn}
\begin{array}{l}
  n_1+\cdots +n_K = N \\
  A_1\cdot n_1+\cdots +A_K\cdot n_K = M\cdot N
\end{array}
\end{equation}
When the number $N$ is large, we may treat
\begin{equation}\label{epk}
    p_k\;=\;\frac{n_k}{N}
\end{equation}
The solution of \eqref{emn} with respect to $\{n_1,\ldots , n_K\}$
is, however, far from being unique. Meanwhile, the solutions do not
possess equal rights: each particular solution $\{n_1,\ldots ,
n_K\}$, according to Bernoulli formula, has \emph{a priori}
probability
\[
P(n_1,\ldots, n_K) = \frac{N!}{n_1!\cdots n_K!}\;p_1^{n_1}\cdots
p_K^{n_K}
\]
Maximizing the value of the probability $P(n_1,\ldots n_K)$, among
the solutions $\{n_1,\ldots , n_K\}$, satisfying \eqref{emn} we find
one, which has greatest probability, therefore we . Using Stirling
formula we get (see, e.g. \cite{kullback} for details):
\begin{equation}\label{elnw}
    \log P(n_1,\ldots n_K)
    \sim
    N\cdot \left(-\frac{n_1}{N}\log\frac{n_1}{N}-\cdots-
    \frac{n_K}{N}\log\frac{n_K}{N}\right)
\end{equation}
The above formula is the Shannon entropy
\begin{equation}\label{eshann}
    \frac{1}{N}\,\log P(n_1,\ldots n_K)
    \sim
    -p_1\log p_1-\cdots-
    p_K \log p_K
\end{equation}
and the maximum of $\log P(n_1,\ldots n_K)$ is attained at
\begin{equation}\label{einientr}
    p_k\;=\;
    \frac{n_k}{N}
    \sim
    \frac{e^{-\beta A_k}}{Z}
\end{equation}
where $Z$ is the normalizing factor
\begin{equation}\label{ez}
    Z=\sum_k e^{-\beta A_k}
\end{equation}
and $\beta$ is a `temperature parameter', obtained by solving
\eqref{emn} in explicit form
\begin{equation}\label{ebeta}
    \frac{A_1 e^{-\beta A_1}+\cdots+A_K e^{-\beta A_K}}{e^{-\beta A_1}+\cdots+e^{-\beta A_K}}
    \;=\;
    \frac{M}{N}
\end{equation}
with respect to $\beta$. This gives us definite values of
$p_1,\ldots p_K$\footnote{Recall that we know only $M$ and we wish
to infer from this knowledge the `natural' values of $p_1,\ldots
p_K$.}.

As an illustration, consider a usual die, that is $K=6$,
$A_1=1,\ldots,A_6=6$. Begin with a symmetric case
$M=(1+2+\cdots+6)/6=3.5$. The solution of \eqref{ebeta} is
$\beta=0$, which means that the Laplace principle still works and
this particular value of $M$ gives no preference to any state,
therefore the null hypothesis (the uniform distribution $p_j=1/6$)
should not be rejected.

If the die is \emph{`biased'}, we obtain a different value o $M$,
say, $M=2.5$. In this case the Laplace principle should be expanded:
namely, we search the distribution maximizing the entropy $H=-\sum
p_j \log p_j$. In our particular case this gives the following
answer:
\begin{equation}\label{edicent}
    p_j
    =
    \frac{e^{-\beta j}}{Z}
\end{equation}
Solving numerically \eqref{ebeta} for $M=2.5$ gives us
$\beta=0.3710$, that is
\[
\{p_1, p_2,\ldots,p_6\}\;=\;\{0.3476, 0.2396, 0.1654, 0.1143,
0.0788, 0.0543\}
\]

The main message of this section is the following. We provide a
completely classical example where we have \emph{no knowledge} about
the input state (distribution) but we \emph{need} to tell something
about it. A principle is described to choose a concrete distribution
on the basis of a given small amount of knowledge.

\section{Continuous ensembles}\label{scontinens}
Why the idea to maximize the entropy $H$ is a development of Laplace
idea of symmetry and non-preference? For any given average value we
consider all possible distributions which yield this average value.
Then the take such distributions which are typical, that is, which
mostly occur in all possible configurations \cite{nhad}. The
preference is given to what occur with maximal number of
combinations, expressed by the statistical weight

\[
P = \frac{N!}{\prod_j n_j} \sim N \cdot \left( -\sum_j\limits
\frac{n_j}{N}\,\log\frac{n_j}{N}\right)
\]
where $N$ is the total number of trials and $n_j$ is number of
occurrence of $j$-th face ($j$-th outcome, more generally).

Now let us develop a similar construction, but passing from numbers
to operators, that is, the mean value is now an operator rather than
a number $M$ in \eqref{emn}. The restriction \eqref{emn} becomes of
matrix form. The consequence of this is that the value of the
parameter $\beta$---appropriate Lagrange multiple---becomes matrix
as well.

Let $\hh=\cfield^\lth$ be an $\lth$-dimensional Hermitian space, let
$\rho$ be a density matrix in $\hh$. There are infinitely many
ensembles whose average density matrix is $\rrh$. Among them we
would like to emphasize a `natural' one. First suppose this ensemble
to be finite and, like in previous section, in order to find a
natural ensemble, maximize its mixing entropy. The result is the
following: given any arbitrary large number $E$, we can always find
an ensemble of $2^E$ pure states which averages to $\rrh$ and whose
mixing entropy is $E$: this is a uniform ensemble \cite{rrz0403105}.
So, there is no limit for mixing entropy for finite ensembles. As a
result, we pass to continuous ensembles with the distribution
density expressed by a function $\mu(\phi)$ where $\phi$ ranges over
all unit vectors\footnote{Pure states form a projective space rather
than the unit sphere in $\hh$. On the other hand, one may integrate
over any probabilistic space. Usually distributions of pure states
over the spectrum of observables are studied, sometimes probability
distributions on the projective spaces are considered \cite{sqprop}.
In this paper for technical reasons we prefer to represent ensembles
of pure states by measures on unit vectors in $\hh$. We use the
Umegaki measure on $\cbn$--- the uniform measure with respect to the
action of $U(\lth)$ normalized so that $\int_{\cbn}\mesb=1$.} in
$\hh$.

The set of all self-adjoint operators in $\hh=\cfield^\lth$ has a
natural structure of a real space $\rfield^{2\lth}$, in which the
set of all density matrices is a hypersurface, which is the zero
surface $T=0$ of the affine functional $T=\trm{}X-1$. The density
operator of a continuous ensemble associated with the measure
$\mu(\phi)$ on the set $\cbn$ of unit vectors in $\hh$ is calculated
as the following (matrix) integral

\begin{equation}\label{e01integral}
  \rrh
  \;=\;
  \int_{\phi\in\cbn}\limits\;
  \mu(\phi)\,
  \raypr{\phi}
  \,\mesb
\end{equation}

\noindent where $\raypr{\phi}$ is the projector onto the vector
$\bra{\phi}$ and $\mesb$ is the above mentioned normalized measure
on $\cbn$, that is,$\int_{\phi\in\cbn}\limits\;
  \,\mesb
\;=\; 1$. Effectively, the operator integral $\rrh$ in
\eqref{e01integral} can be calculated by its matrix elements. In any
fixed basis $\{\ket{\ee_i}\}$ in $\hh$, each its matrix element
$\rrh_{ij}=\bracket{\ee_i}{\rrh}{\ee_j}$ is the following numerical
integral:

\begin{equation}\label{e01basis}
\rrh_{ij} \;=\;
  \bracket{\ee_i}{\rrh}{\ee_j}
  \;=\;
  \int_{\phi\in\cbn}\limits\;
  \mu(\phi)\,
  \braket{\ee_i}{\phi}
  \braket{\phi}{\ee_j}
  \,\mesb
\end{equation}

\paragraph{Kullback--Leibler distance.}
We quantify the state preparation efforts by the difference between
the entropy of uniform distribution (that is, our null hypothesis)
and the entropy of the ensemble\footnote{We are speaking here of
\emph{mixing entropy} \cite{wehrl} of the ensemble rather than about
von Neumann entropy of its density matrix.} in question. This is
equal to Kullback-Leibler distance \cite{kullback}
\[\entrp(\dst\|\dst_0) \;=\;
\int \dst(x)\ln\frac{\dst(x)}{\dst_0(x)} \rmd x
\] between the distribution $\dst(x)$ and the uniform distribution
$\dst_0(x)$ with constant density, normalize the counting measure
$\rmd x$ on the probability space so that $\dst_0=1$. This distance
is the average likelihood ratio, on which the choice of statistical
hypothesis is based. Then, in order to minimize the Type~I error we
have to choose a hypothesis with the smallest average likelihood
ratio.

\paragraph{Maximizing the entropy.} The problem reduces to the following. For given density
matrix $\rrh$ find a continuous ensemble $\mu$ having minimal
differential entropy:
\begin{equation}\label{egenprob}
\entrp(\dst)\;=\; \int \dst(x)\ln\dst(x) \rmd x \;\to\; \min,\qquad
\int \prjo{\psi}\, \dst(\psi) \mesb \;=\; \rrh
\end{equation}
where $\mesb$ is the unitary invariant measure on pure states
normalized to integrate to unity. When there is no constraints in
\eqref{egenprob}, the answer is straightforward---the minimum (equal
to zero) is attained on uniform distribution---this situation is
quite similar to the symmetric classical case considered in section
\ref{sclass}. To solve the problem with constraints, we use the
Lagrange multiples method \cite{lazygoga}. The appropriate Lagrange
function reads:
\[
\mathcal{L}(\dst) \;=\; \entrp(\dst)\;-\; \trm\,\lmu\left(\int
\prjo{\psi}\, \dst(\psi) \mesb \;-\; \rrh\right)
\]
where the Lagrange multiple $\lmu$ is a matrix since the constraints
in \eqref{egenprob} are of matrix character. Substituting the
expression for $\entrp(\dst)$ and making the derivative of
$\mathcal{L}$ over $\dst$ zero, we get
\begin{equation}\label{egibbs}
\dst(\psi) \;=\;
\frac{\rme^{-\,\trm\tpar\prjo{\psi}}}{Z\left(\tpar\right)}
\end{equation}
where $\tpar$ is the optimal value of the Lagrange multiple $\lmu$
which we derive from the constraint \eqref{egenprob} and the
normalizing multiple
\begin{equation}\label{epartfun}
Z(\tpar) \;=\; \int \rme^{-\,\trm\tpar\prjo{\psi}} \mesb
\end{equation}
is the partition function for \eqref{egibbs}.

\section{Conclusions}

In Classical Mechanics, there is a unique correspondence between mixed states and ensembles of pure states. This is no longer the case in quantum mechanics: if we are give a state described by a density matrix $\rrh$, there are infinitely many ensembles of
pure quantum states, which average to $\rrh$. In our paper we consider quantum systems with finite-dimensional state space $\hh=\cfield^n$. An ensemble of pure states is understood
in a mostly general sense as certain distribution of pure states in H, rather than a finite weighted sum.

Contrary to conventional approach, we exploit continuous distributions of state vectors (but still in finite dimensions $\cfield^n$). The task we tackle is the following, Suppose
we are given a quantum state with a density matrix $\rrh$, and this is all we know about the preparation procedure. In this setting, what could we say about the ensemble, which gave rise to $\rrh$? In order to answer this question, we use standard statistical approach: among all ensembles averaging to $\rrh$ we choose the one which is more spread than the others. What means 'more spread'?

We consider all ensembles averaging to $\rrh$ and, according to Laplace, give no preference to any of them. As stated above, by 'ensemble' we mean a distribution and as a zero hypothesis we take it to be uniform – no knowledge, no preference. However, when
$\rrh\neq 1$, this hypothesis is not compatible with our knowledge that the average quantum state is $\rrh$. In order to comply with Laplace principle \cite{nhad} we choose the distribution of
pure states which 
\begin{itemize}
  \item averages to the state $\rrh$
  \item has the greatest differential entropy
\end{itemize}

\noindent The resulting distribution has the form \eqref{egenprob}: $S(\mu) = \int \mu(\phi) \ln \mu(\phi) \drm\phi \rightarrow \min$. So, summarizing our result 
\begin{quote}
   if we have a source of particles whose average quantum state is $\rrh$, and this is the only information about the source of this particles, we have to state that they are prepared as follows: pure states are emitted with probability density 
\[
\int\;\rme^{-\bracket{\phi}{B}{\phi}}\;\ketbra{\phi}{\phi}\;\drm\phi
\]
\end{quote}

\paragraph{Acknowledgements.} The second author acknowledges the hospitality of organizers and
particularly Jaros\l aw Pykacz. The work was carried out under the
auspices of the research grant RFFI 0706-00119.

\newpage

\appendix

\section*{The existence of lazy ensembles}
Let $\rrh$ be a nondegenerate density operator, that is,
$0<\lambda_1\leq\cdots\leq\lambda_n$, where $\{\lambda_j\}$ are the
eigenvalues of the operator $\rrh$. We are going to prove that for
any such $\rrh$ there exists an operator $Y$ such that
\begin{equation}\label{emaineq}
\int\;\rme^{\bracket{\phi}{Y}{\phi}}\;\ketbra{\phi}{\phi}\;\drm\phi
\;=\;\rrh
\end{equation}
From now on all the integrals are taken over the unit sphere in
$\cfield^n$ (if otherwise not written explicitly), and $\drm\phi$ is
the invariant measure on it induced by Lebesgue measure in
$\rfield^{2n}$. Fix a density matrix $\rrh$ and consider the
scalar-valued function
\begin{equation}\label{e112a}
Z_{\infty}
\;=\;
\int\;\rme^{\bracket{\phi}{Y}{\phi}}\;\drm\phi
-
\trm (Y\rrh)
\end{equation}
and search for its minimum. If the minimum exists, then then the
necessary condition for it is vanishing of all partial derivatives
\[
\frac{\partial Z_{\infty}}{\partial Y}
\;=\;0
\]
which is equivalent to \eqref{emaineq}, that is, if the minimal
value of \eqref{e112a} exists, then $Y$, at which it is attained,
yields the solution for \eqref{emaineq}. So, we only have to prove the
existence of the minimum of the function $Z_{\infty}$ \eqref{e112a}.
Consider a sequence of functions $Z_M$ parameterized by integer $M$:
\[
Z_M(Y)
\;=\;
\int\left(
1+
\frac{\bracket{\phi}{Y}{\phi}}{2M}
\right)^{2M}\drm\phi
\,-\,\trm (Y\rrh)
\]
for them, the existence of minimum is straightforward since $Z_M$ is
a mixture of positive concave functions. That means, the (operator)
equation
\begin{equation}\label{epolyop}
\int\left(
1+
\frac{\bracket{\phi}{Y}{\phi}}{2M}
\right)^{2M-1}\ketbra{\phi}{\phi}\drm\phi
\;=\;\rrh
\end{equation}
always has a solution $Y$ (this $Y$ depends on $M$). Let us evaluate
the eigenvalues of $Y$, denote them $y_1\le\ldots\le y_n$
appropriately ordered. That is, for any $\phi$
\[
y_1\le\bracket{\phi}{Y}{\phi}\le y_n
\]
hence
\[
\left(
1+
\frac{y_1}{2M}
\right)^{2M-1}\int\ketbra{\phi}{\phi}\drm\phi
\le\]\[\le
\int\left(
1+
\frac{\bracket{\phi}{Y}{\phi}}{2M}
\right)^{2M-1}\ketbra{\phi}{\phi}\drm\phi
\le\]\[\le
\left(
1+
\frac{y_n}{2M}
\right)^{2M-1}\int\ketbra{\phi}{\phi}\drm\phi
\]
taking the trace we get
\[
\left(
1+
\frac{y_1}{2M}
\right)^{2M-1}\le
\trm\rrh\le
\left(
1+
\frac{y_n}{2M}
\right)^{2M-1}
\]
Since we are dealing with a density operator $\rrh$, its trace
equals 1, therefore
\begin{equation}\label{evalbsimple}
    y_1\le 0 \le y_n
\end{equation}

\medskip

Now let us evaluate the difference $y_n-y_1$ between the greatest
and the least eigenvalues of $Y$ from \eqref{epolyop}. Let
$\ket{\ee_j}$ be an eigenvector of $\rrh$ associated with the
eigenvalue $\lambda_j$, then \eqref{epolyop} reads:
\[
\lambda_j
\;=\;
\hspace{-2em}\int_{t_1+\cdots+t_{n-1}\le 1}\limits
\left( 1+ \frac{y_1 t_1
+\cdots+ y_{n-1} t_{n-1} + y_n(1-t_1-\cdots-t_{n-1})}{2M}
\right)^{2M-1} t_j \drm t_1\cdots \drm t_{n-1}
\]
therefore \((y_n-y_1)\lambda_1=\)
\[
\;=\;
\hspace{-2em}\int_{t_2+\cdots+t_{n-1}\le 1}\limits
\hspace{-2em}\drm t_2 \cdots \drm t_{n-1}
\hspace{-1em}\int_{t_1=0}^{1-t_2-\cdot-t_{n-1}}\limits
\left(
1+\frac{(y_1-y_n) t_1+y_n + \sum_{k=2}^{n-1} (y_{k}-y_n) t_{k}}{2M}
\right)^{2M-1}
\hspace{-1em}t_1 (y_n-y_1) \drm t_1
\;=
\]
\[
\;=\;
-\hspace{-2em}
\int_{t_2+\cdots+t_{n-1}\le 1}\limits
\hspace{-2em}\drm t_2 \cdots \drm t_{n-1}
\left[
\left(
1+\frac{(y_1-y_n) t_1+y_n + \sum_{k=2}^{n-1} (y_{k}-y_n) t_{k}}{2M}
\right)^{2M}t_1
\right|_{\,t_1=0}^{\,1-t_2-\cdot-t_{n-1}}
-\]
\[-
\left.
\hspace{-2em}\int_{t_1=0}^{1-t_2-\cdot-t_{n-1}}\limits \left(
1+\frac{(y_1-y_n) t_1+y_n + \sum_{k=2}^{n-1} (y_{k}-y_n) t_{k}}{2M}
\right)^{2M}\drm t_1
\right]
 \]
The first summand in the above expression is minus an integral of a non-negative function, the second is the following integral over the unit sphere 
\[
\int_{t_1+\cdots+t_{n-1}\le 1}\limits
\left(
1+\frac{(y_1-y_n) t_1+y_n + \sum_{k=2}^{n-1} (y_{k}-y_n) t_{k}}{2M}
\right)^{2M}\drm t_1\drm t_2 \cdots \drm t_{n-1}
\]
denote it by $K$ and rewrite in a more familiar form: 
\[
K
\;=\;
\int_{t_1+\cdots+t_{n}= 1}\limits
\left(
1+\frac{\sum_{k=1}^{n} y_{k} t_{k}}{2M}
\right)^{2M}\drm t_1\drm t_2 \cdots \drm t_{n-1}
\;=\;
\int\left(
1+\frac{\bracket{\phi}{Y}{\phi}}{2M}
\right)^{2M}\drm \phi
\]
then 
\begin{equation}\label{eevaldk}
    (y_n-y_1)\lambda_1
    \;\le\;
    K
\end{equation}
Using \eqref{epolyop} and taking into account that $\left(
1+\frac{\bracket{\phi}{Y}{\phi}}{2M}
\right)^{2M-1}\left(
1+\frac{\bracket{\phi}{Y}{\phi}}{2M}
\right)$, we have
\[
K
\;=\;
\trm \rrh
\;+\;
\int
\left(
1+\frac{\bracket{\phi}{Y}{\phi}}{2M}
\right)^{2M-1}
\frac{\bracket{\phi}{Y}{\phi}}{2M}
\drm \phi
\;\le\;
\]
\[
\;\le\;
\trm \rrh
\;+\;
\trm \rrh
\frac{y_n-y_1}{2M}
\;=\;
1+\frac{y_n-y_1}{2M}
\]
since $y_n-y_1\le\bracket{\phi}{Y}{\phi}$. Therefore
\[
(y_n-y_1)\lambda_1
    \;\le\;
    K\;\le\;
    1+\frac{y_n-y_1}{2M}
\]
so $(y_n-y_1)\left(\lambda_1-\frac{1}{2M}\right)\;\le\;1$. That is, for any $M>\lambda_1$ we have $\lambda_1-\frac{1}{2M}\ge\frac{\lambda_1}{2}$, therefore
\begin{equation}\label{efinaldiff}
    (y_n-y_1)\le 2/\lambda_1
\end{equation}

\medskip

This means that for sufficiently big $M$ the solutions $\{y_1,\ldots,y_n\}$ of \eqref{emaineq} remain in the compact set:
\[
\left\lbrace
\begin{array}{l}
  y_1\le 0 \le y_n \\
  (y_n-y_1)\le 2/\lambda_1
\end{array}
\right.
\]
Therefore the limit of the solutions exist which means that for any nondegenerate density operator $\rrh$ there always exists the appropriate lazy ensemble
\[
\int\;\rme^{-\bracket{\phi}{B}{\phi}}\;\ketbra{\phi}{\phi}\;\drm\phi
\;=\;\rrh
\]
\end{document}